\documentclass[floatfix,showpacs,prl,aps]{revtex4}
\usepackage[applemac]{inputenc}
\usepackage[T1]{fontenc} 
\usepackage{float}
\usepackage{amssymb}
\usepackage{enumerate}
\usepackage{braket}
 \usepackage{amsmath} 
\usepackage{subfigure}
\usepackage{lipsum}
\usepackage{amsfonts} 
\usepackage{amssymb, mathrsfs}
\usepackage{amsmath}
\usepackage{bm}
\usepackage[pdftex]{graphicx}
\usepackage{footnote}
\usepackage[colorlinks]{hyperref}
\usepackage{titlesec}
\usepackage{graphicx} 
\usepackage{tikz}
\usepackage[compat=1.1.0]{tikz-feynman}

\def\beq{\begin{equation}}
\def\eeq{\end{equation}}
\def\bsp{\begin{split}}
\def\esp{\end{split}}
\def\bea{\begin{eqnarray}}
\def\eea{\end{eqnarray}}
\def\ba{\begin{array}}
\def\ea{\end{array}}

\def\haf#1{{{#1}\over 2}}

\def\lb{\left(}
\def\rb{\right)}
\def\lr{\left|}
\def\rr{\right|}
\def\lbr{\left\{}
\def\rbr{\right\}}
\def\l.{\left.}
\def\r.{\right.}

\def\ie{{\it i.e. }}

\def\part{\partial}

\newcounter{subsections}

\titleformat{\subsection}
  {}
  {0pt}
  {}                    
  {\bfseries\addtocounter{subsections}{1}\thesubsections. } 

\begin{document}

\preprint{UdeM-GPP-TH-21-xxx}
\preprint{arXiv:21xx.xxx}
\title{What is the Gravitational Field of a Mass in a Spatially Nonlocal Quantum Superposition?}
\author{Rémi Ligez$^{1}$} 
\email{remi.ligez@umontreal.ca}
\author{R. B. MacKenzie$^{1}$} 
\email{richard.mackenzie@umontreal.ca}
\author{Victor Massart$^{1}$} 
\email{victor.massart@umontreal.ca}
\author{M. B. Paranjape$^{1,2}$} 
\email{paranj@lps.umontreal.ca}
\author{U. A. Yajnik$^{3}$} 
\email{yajnik@iitb.ac.in}

\affiliation{$^{1}$Groupe de physique des particules, D\'epartement de physique,
Universit\'e de Montr\'eal, C.P. 6128, succ. centre-ville, Montr\'eal, 
Qu\'ebec, Canada, H3C 3J7 }
\affiliation{$^{2}$Centre de recherche math\'ematiques, Universit\'e de Montr\'eal, C.P. 6128, succ. centre-ville, Montr\'eal, Qu\'ebec, Canada, H3C 3J7 }
\affiliation{$^3$  Department of Physics, Indian Institute of Technology Bombay, Powai, Mumbaï - 400076}

\vspace{0.8cm}

\begin{abstract}
The study of the gravitational field produced by a spatially non-local,  superposed quantum state of a massive particle is a thrilling area of modern physics.  One question to be answered is whether the gravitational field behaves as the classical superposition of two particles separated by a spatial distance with half the mass located at each position or as a quantum superposition with a far more interesting and subtle behaviour for the gravitational field.  Quantum field theory is ideally suited to probe exactly this kind of question.  We study the scattering of a massless scalar on such a spatially nonlocal, quantum superposition of a massive particle.  We compute the differential scattering cross section corresponding to the interaction coming from the exchange of one graviton.  We find that the scattering cross section is not at all represented by the Schrödinger-Newton picture of potential scattering from two localized sources with half the mass at each source.  We discuss how our result would be lethal to the Schrödinger-Newton description of gravitation interacting with quantum matter and would be conducive to considering the gravitational field to be quantized.   We comment on the experimental feasibility of observing such effects.
\end{abstract}


\maketitle

\section{Introduction}

At the core of modern physics lie the two most successful theory of physics : General Relativity (GR) on one side and Quantum Field Theory (QFT) on the other. Consistently combining these theories is one of the main goals of modern theoretical physics. The most common approach is to consider the quantum theory to be more fundamental and try to quantize gravity \cite{Kiefer}. This choice is also justified by some indirect evidence \cite{PageGilkerIndirectEvidence}. The opposite direction has also been advocated, of ``gravitizing'' quantum mechanics, with some interesting arguments, see Penrose \cite{Penrose2014gravitization, CarlipIsQuantum}.

However, we shall be interested in the standard ideas of a quantum theory of gravity.  Indeed, quantum gravity as a low energy effective field theory makes perfect sense \cite{,Donoghue-prl,PhysRevD.50.3874,Donoghue:1995cz,Burgess:2003jk}, although the directly observable predictions of such an effective quantum field theory are still not remotely experimentally observable.   An intriguing possibility for the detection of noise due to quantum fluctuations in the gravitational field at gravitational wave detectors such as LIGO \cite{aLIGO} has been suggested recently \cite{Parikh}.   Within the effective field theory point of view, gravitation must be treated as a fully quantum mechanical field with all the complexities ensuing from the possibilities of arbitrary linear superpositions of quantum states.   It is in this context that we examine the gravitational scattering of a massless particle from the effective quantum gravitational field of a massive particle in a spatially non-local quantum superposition.

\section{Scattering on a spatially nonlocal wave packet}

The usual formalism for scattering in quantum field theory corresponds to computing the amplitude for a given scattering process and converting the amplitude into an experimentally observable differential scattering cross section.  The formalism is almost as old as the advent of quantum field theory, however, generally, certain assumptions are made so that the dependence on the initial and final wave packets disappears; see for example \cite{Peskin:1995ev}.  However, here we wish to consider a process where the cross section strongly depends on the wave packet of one of the initial states, a situation which is not usually considered.  

We consider as the scatterer a particle of mass $M$ in a linearly superposed state described by a spatially non-local wave function, which of course correspond to a particular momentum space wavefunction $\phi_1(\bm k_1)$, causing the scattering of a massless particle supposed to be in a state $\phi_2(\bm k_2)$.  While $\phi_2(\bm k_2)$ is assumed to be sharply peaked at momentum $\bm p_2$,  the wavefunction $\phi_1(\bm k_1)$  is not peaked at any $\bm p_1$, but nevertheless assumed for simplicity to be symmetrically distributed around  $\bm p_1=0$, \ie $\phi_1(-\bm k_1)=\phi_1(\bm k_1)$.   This means that the calculation is done in the convenient reference frame where $\bm p_1=0$, i.e., the ``centre of mass'' frame of the wavefunction.

The formula for the differential of the cross section is given by, :
\bea
d\sigma&=\lb \prod_f\frac{d^3p_f}{(2\pi)^3}\frac{1}{2E_f}\rb\displaystyle\int\frac{d^3k_{1}}{(2\pi)^3}\displaystyle\int\frac{d^3k_{2}}{(2\pi)^3}\frac{\lr{\cal M}\lb k_{1},k_{2}\to \lbr p_f\rbr\rb\rr^2}{2E_{1}2E_{2}\lr v_{1}-v_{2}\rr}\\
&\times \lr \phi_{1}\lb \bm k_{1}\rb\rr^2\lr \phi_{2}\lb \bm k_{2}\rb\rr^2(2\pi)^4\delta^4\lb k_{1}+k_{2}-\sum p_f\rb .
\eea
For our needs, we will use the formalism of Kotkin et al \cite{kotkin} as elaborated in Karlovets \cite{Karlovets} that employs the Wigner function formalism \cite{Wigner}. We will take the incident massless particles in on-shell momentum space wave packets $\phi_{2}(\bm  k_2)$ centred and highly peaked on a momentum $\bm  p_{2}$.  These massless particles will be incident on particles of mass $M$  which are in on-shell, spatially nonlocal wave packets. When expressed in terms of momentum space wave packets $\phi_{1}(\bm  k_1)$, these wave packets are not highly peaked on any specific momentum as noted above.     We will be interested in the inclusive scattering cross section for the massless particle, $p_2\to p_4$, while the non-trivial wave packet will in principle scatter to all possible allowed final states, which will be integrated over.  

It is most convenient to consider the wave function of the scattered massive particle, that was initially in the spatially nonlocal wave function, to be scattered to wave functions that are eigenstates of momentum $\bm  p_3$ and to integrate over this momentum, as these do correspond to a complete set of final states.  In practice, the integration is of course not required as energy-momentum conservation for given $k_1, p_2, p_4$ fixes the value of $p_3$. Thus the scattering will give rise to final momenta  $k_1\to p_3$ and  $p_2\to p_4$ with integration over $\bm  k_1$ smeared with wave function $\phi_{1}(\bm  k_1)$ understood.   Then straightforwardly simplifying the formula in \cite{Karlovets}, to the case of a particle in a momentum eigenstate scattering on a particle in a non-trivial wave packet, we find
\begin{equation}
    d\sigma = \int \frac{d^3k_1}{(2\pi)^3}  \vert T_{PW}  \lb\lbr k_1,p_2\rbr\to \lbr p_3,p_4\rbr\rb \vert^2  \vert \phi_1 (\bm  k_1) \vert^2 (2\pi)^4\delta^{(4)} (k_1 +  p_2  -p_3 - p_4) \frac{d^3 p_3}{(2\pi)^3} \frac{d^3 p_4}{(2\pi)^3}\label{1}
\end{equation}
where the plane wave amplitude $T_{PW} \lb\lbr k_1,p_2\rbr\to \lbr p_3,p_4\rbr\rb$ is given by
\begin{equation}
    T_{PW} \lb\lbr k_1,p_2\rbr\to \lbr p_3,p_4\rbr\rb = \frac{\mathcal{M} \lb\lbr k_1,p_2\rbr\to \lbr p_3,p_4\rbr\rb}{\sqrt{2\epsilon_1 2\epsilon_2 2\epsilon_3 2\epsilon_4}}
\end{equation}
with ${\cal M} \lb\lbr k_1,p_2\rbr\to \lbr p_3,p_4\rbr\rb$  the invariant matrix element for the momentum space scattering process \cite{Peskin:1995ev}, and $\epsilon_i$ the energy of particle $i$.  We note the usual factor of ${\lr v_1-v_2\rr}$ in the particle flux simplifies to unity, the velocity of the massless particle, because of the assumed symmetry of the wave function $\phi_1(\bm  k_1)$.

The normalized wave function for the particle of type 1, which is taken to be spatially nonlocal, in momentum space has the form
\begin{equation}
\vert \phi_1 (\bm  k_1) \vert^2 = 4 \left( \pi \sigma^2 \right)^{3/2} e^{- \sigma^2 \vert \bm  k_1 \vert^2} \frac{2 + e^{2i \bm  r \cdot \bm  k_1} + e^{-2i \bm  r \cdot \bm  k_1}}{1 + e^{-  \vert \bm  r\vert^2/(\sigma^2)}}
\end{equation}
where $\sigma$ is the width of a Gaussian wave packet that is superposed at spatial position $\bm  r$ and $\bm {-r}$.

\section{The 1-graviton exchange scattering amplitude}
The amplitude is easily computed using the linearized gravitational theory and subsequent graviton propagator and matter vertices, following Donoghue \cite{PhysRevD.50.3874,Donoghue:1995cz} (see also \cite{DeWittI,DeWittII,DeWittIII,veltman1976quantum,Burgess:2003jk,Scadron}), as prescribed by the Feynman diagram Fig. (\ref{fd})
\begin{center}
\begin{figure}
\begin{tikzpicture}
  \begin{feynman}
    \vertex (b);
    \vertex [above left=of b] (f1) {\( p_3 \)};
    \vertex [below left=of b] (a) {\(\phi_1, k_1 \)};
    \vertex [right=of b] (c);
    \vertex [above right=of c] (f2) {\( p_4 \)};
    \vertex [below right=of c] (f3) {\( p_2 \)};
 
    \diagram* {
      (a) -- [fermion] (b),
      (b) -- [fermion] (f1),
      (b) -- [gluon, edge label'=\(q\)] (c),
      (c) -- [fermion] (f2),
      (f3) -- [fermion] (c),
    };
  \end{feynman}
\end{tikzpicture}
\caption{1-graviton exchange scattering\label{fd}}
\end{figure}
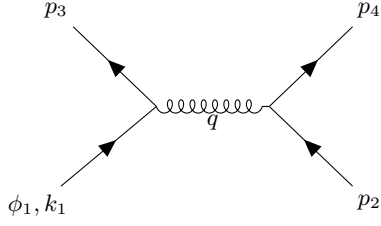
\end{center}
Applying momentum conservation, we get for the amplitude
\begin{align}
    \mathcal{M} &= \frac{(-i)^2\kappa^2 (k_1^\mu p_3^\nu + k_1^\nu p_3^\mu - (p_2 \cdot p_4) \eta^{\mu \nu} ) (\eta_{\mu \alpha}  \eta_{\nu \beta}  + \eta_{\mu \beta}  \eta_{\nu \alpha}  - \eta_{\mu \nu}  \eta_{\alpha \beta} ) (p_2^\alpha p_4^\beta + p_2^\beta p_4^\alpha - (p_2 \cdot p_4) \eta^{\alpha \beta} )}{2q^2} \nonumber \\
    &=  \frac{\kappa^2\left[ 2 (k_1 \cdot p_2) (k_1 \cdot p_4) + (k_1 \cdot p_2 )(p_2 \cdot p_4)- (k_1 \cdot p_4)(p_2 \cdot p_4)  -  (p_2 \cdot p_4)^2\right]}{ p_2 \cdot p_4} \\
     &= \frac{\kappa^2}{1-\bm  \hat p_2 \cdot \bm  \hat p_4} \Big[2 M^2  -2M \bm  k_1 \cdot (\bm  \hat p_2 + \bm  \hat p_4)+ M (\omega_2 - \omega_4) (1 - \bm  \hat p_2 \cdot \bm  \hat p_4) + 2\vert \bm  k_1 \vert^2  \nonumber \\
     &+  2 (\bm  k_1 \cdot \bm  \hat p_2) (\bm  k_1 \cdot \bm  \hat p_4) - \bm  k_1 \cdot (\omega_2 \bm  \hat p_2 - \omega_4 \bm  \hat p_4) (1 - \bm  \hat p_2 \cdot \bm  \hat p_4) - \omega_2 \omega_4 (1 - \bm  \hat p_2 \cdot \bm  \hat p_4)^2 \Big]
\end{align}
where $\kappa$ is the gravitational coupling constant, $\omega_2$ and $\omega_4$ are short hand notation for $p_2^0$ and $p_4^0$ and the momentum transfer is $q=p_3-k_1=p_2-p_4$ with $q^2=-2p_2\cdot p_4=-2\omega_2\omega_4(1-\bm  {\hat p_2\cdot\bm  \hat p_4})$ as $p_2$ and $p_4$ are massless and on shell.
\section{The scattering cross section}
The amplitude must now be folded in with the wave function, the energy denominators and factors of $2\pi $ as in Eqn. \eqref{1} and then integrated over $d^3p_3$ which removes the spatial delta functions and integrated over $d\omega_4$ which removes the temporal delta function yielding the differential scattering cross section 
\beq
\frac{d\sigma}{d\Omega_4}=\int \frac{d^3k_1}{2^9\pi^5} \vert \phi_1 (\bm k_1) \vert^2 \frac{\lr\mathcal{M}\rr^2}{{\epsilon_1 \omega_2  \epsilon_3 \omega_4}}\frac{\omega_4^2}{\lr f'_\delta(\omega_4)\rr}.\label{6}
\eeq
The energy conserving delta function is given by $\delta(f_\delta(\omega_4))=\delta(\epsilon_1+\omega_2-\epsilon_3-\omega_4)$ where the complications arise because $\epsilon_3=\sqrt{M^2 +({\bm  k_1 +(\bm p_2} -\omega_4   (\bm {\hat p_4})^2}$.  $\epsilon_1=\sqrt{M^2+\bm  k_1^2}$ is the energy of the Fourier component corresponding to momentum  $\bm  k_1$ of the spatially nonlocal particle wave function and $\omega_2$ is the energy of the incoming massless particle.   The full expression for the cross section is a complicated, unenlightening jumble, however its multipole expansion does shed some light on the gravitational interaction.  

\section{Multipole expansion of the scattering cross section}
We can write the cross section Eqn.\eqref{6} as
\beq
\frac{d\sigma}{d\Omega_4}=\int \frac{d^3k_1}{2^9\pi^5} \vert \phi_1 (\bm k_1) \vert^2 g(\bm  k_1)\lr\mathcal M\rr^2
\eeq
and then $g(\bm  k_1)\lr\mathcal M\rr^2$ admits an expansion in powers of $\bm  k_1$ as
\beq
g(\bm  k_1)\lr\mathcal M\rr^2=\l. g(\bm  k_1)\lr\mathcal M\rr^2\rr_{\bm  k_1=0}+\haf 1\l.\partial_{\bm  k_1^i}\partial_{\bm  k_1^j}\lb g(\bm  k_1)\lr\mathcal M\rr^2\rb\rr_{\bm  k_1=0}{\bm  k_1^i}{\bm  k_1^j} +\cdots
\eeq
where the terms odd in $\bm  k_1$ are absent due to symmetry.   Then the scattering cross section admits the multipole expansion
\begin{align}
   \frac{d\sigma}{d\Omega_4}= & \alpha \int \vert \phi_1 (\bm  k_1) \vert^2 \frac{d^3k_1}{(2\pi)^3} + \beta^{ij} \int k_{1i} k_{1j} \vert \phi_1 (\bm  k_1) \vert^2 \frac{d^3k_1}{(2\pi)^3} \\
    &= \alpha  + \frac{\beta^{ij} \delta_{ij}}{3} \int \vert \bm  k_1\vert^2 \vert \phi_1 (\bm  k_1) \vert^2 \frac{d^3k_1}{(2\pi)^3}+ \beta^{ij} \int \left(k_{1i} k_{1j} - \frac{\vert \bm  k_1 \vert^2}{3} \delta_{ij} \right) \vert \phi_1 (\bm  k_1) \vert^2 \frac{d^3k_1}{(2\pi)^3}+\cdots\\
    &=\alpha +\frac{\beta^{ij} \delta_{ij}}{3} \lb\frac{3}{2 \sigma^2} - \frac{\vert\bm  r\vert^2}{\sigma^4} \frac{1}{1 + e^{\vert \bm  r \vert^2/\sigma^2}}  \rb     +  \beta^{ij}\left(\frac{\vert \bm  r \vert^2}{3} \delta_{ij}-  r_{i}   r_{j}  \right)\frac{1}{\sigma^4\lb 1 + e^{\vert \bm  r \vert^2/\sigma^2}\rb}\label{11}
\end{align}
where evidently $\alpha=\l. \frac{1}{2^6\pi^2}g(\bm  k_1)\lr\mathcal M\rr^2\rr_{\bm  k_1=0}$ and $\beta^{ij}=\frac{1}{2^7\pi^2}\l.\partial_{\bm  k_1^i}\partial_{\bm  k_1^j}\lb g(\bm  k_1)\lr\mathcal M\rr^2\rb\rr_{\bm  k_1=0}$ and where we have used the integral
\beq
\int k_{1i} k_{1j} \vert \phi_1 (\bm  k_1) \vert^2 \frac{d^3 k_1}{(2\pi)^3} = \frac{\delta_{ij}}{2 \sigma^2} - \frac{r_i r_j}{\sigma^4} \frac{1}{1 + e^{\vert \bm  r \vert^2/\sigma^2}}.
\eeq

The leading terms in the expansion about $\bm  k_1=0$ are found after a somewhat tedious calculation.  For the second derivative we will use:
\bea
    \l.\partial_{\bm  k_1^i}\partial_{\bm  k_1^j}\lb g(\bm  k_1)\lr\mathcal M\rr^2\rb\rr_{\bm  k_1=0}&=&\nonumber\\
  \mathcal{M}^2 \partial_{\bm  k_1^i}\partial_{\bm  k_1^j} g + 2 \mathcal{M} (\partial_{\bm  k_1^i} g\partial_{\bm  k_1^j}\mathcal{M} &+& \partial_{\bm  k_1^j} g \partial_{\bm  k_1^i} \mathcal{M}) + 2 g(\partial_{\bm  k_1^i}\mathcal{M} \partial_{\bm  k_1^j}\mathcal{M} + \left. \mathcal{M}\partial_{\bm  k_1^i} \partial_{\bm  k_1^j}\mathcal{M}) \right|_{\bm  k_1 = 0}
\eea
We will need the expression for $\omega_4$ obtained from the energy conservation delta function
$\omega_4=\epsilon_1 +\omega_2 -\epsilon_3$, and its derivatives.  We find that the leading terms in the expansion in powers of $1/M$ of $\omega_4$ and its derivatives are given by
\begin{align}
\omega_4 &= \omega_2 - \frac{1}{M} \left( {\omega_2^2} + \bm  k_1 \cdot \bm  p_2- \omega_2 (\bm  k_1 + \bm  p_2)\cdot\bm{\bm {\bm {\bm {\hat p_4 }}}}\right)+ \nonumber\\
    + \frac{1}{M^2} \left( \omega_2^3 - 2\omega_2^2 (\bm  k_1 + \bm  p_2)\cdot\bm{  \bm {\bm {\bm {\hat p_4 }}}}\right. &\left. + \omega_2 (\bm  k_1 \cdot \bm{  \bm {\bm {\bm {\hat p_4 }}}}+ \bm  p_2 \cdot \bm  {\hat p_4})^2 +  (\bm  k_1 \cdot \bm  p_2) (\omega_2 - \bm  k_1 \cdot \bm{  \bm {\bm {\bm {\hat p_4 }}}}- \bm  p_2 \cdot \bm{  \bm {\bm {\bm {\hat p_4 }}}})  \right)  \\
\left. \omega_4 \right|_{\bm  k_1 = 0} &=  \omega_2 \\
 \partial_i \left. \omega_4 \right|_{\bm  k_1 = 0} & =  \frac{\omega_2}{M} (   {\hat p_{4i}}-   {\hat p_{2i}}) \\
 \partial_i \partial_j \left. \omega_4 \right|_{\bm  k_1 = 0} & = \frac{\omega_2}{M^2} (2  {{\hat p_{4i}}{  {\hat p_{4j}}}} -   {\hat p_{2i}}   {\hat p_{4j} }-   {\hat p_{2j} }  {\hat p_{4i} }).
\end{align}
Then we find, using the above, relatively easily, to leading order in powers of $M$
\begin{align}
   \left. \mathcal{M} \right|_{\bm  k_1 = 0} &=  \kappa^2 \frac{2 M^2}{1 - \bm  {\hat p_2 \cdot \bm  \hat p_4} }\\
   \left. \partial_i \mathcal{M} \right|_{\bm  k_1 = 0} &= \frac{\kappa^2}{1 - \bm  {{\hat p_2 \cdot \bm  \hat p_4}}} \left(-2M (  {\hat p_{2i} +   \hat p_{4i}}) \right)\\
   \left. \partial_i \partial_j \mathcal{M} \right|_{\bm  k_1 = 0} &= \frac{\kappa^2}{1 - {\bm  {\hat p_2 \cdot \bm  \hat p_4} }}\left( 4\delta_{ij} + 2 (  {\hat p_{2i}   \hat p_{4j} }+ {  \hat p_{2j}   \hat p_{4i}}) \right).
\end{align}
A lengthier and more tedious calculation, which includes the calculation of $1/|f'_\delta|$, gives $g$ and its derivatives, again expanded in powers of $1/M$,
\begin{align}
    g(\omega_4, \bm  k_1) &= \frac{\omega_4}{ M^2 \omega_2} \Big( 1 - \frac{\omega_2 (1 - \bm { \hat p_2 \cdot \bm  \hat p_4}) - \bm  k_1 \cdot \bm  {\hat p_4}}{M} + \frac{\omega_2^2 - \vert \bm  k_1 \vert^2 - 2 \omega_2 (\bm  k_1 + \bm  p_2)\cdot\bm { {\bm {\bm {\bm {\hat p_4 }}}}}+ (\bm  k_1\cdot\bm  {\bm {\bm {\bm {\hat p_4 }}}+ \bm  p_2 \cdot\bm  \hat p_4})^2}{M^2} \Big) \nonumber \\
    \left. g \right|_{\bm  k_1 =0} &= \frac{1}{ M^2}\\
    \left. \partial_i g \right|_{\bm  k_1 =0} &= \frac{1}{ M^3} (2  {\hat p_{4i} -   \hat p_{2i}}) \\
    \left. \partial_i \partial_j g \right|_{\bm  k_1 =0} &=  \frac{2}{ M^4} \left(3  {\hat p_{4i} \hat p_{4j} -   \hat p_{2i}   \hat p_{4j} -   \hat p_{2j} \hat p_{4i}} - \delta_{ij} \right)
\end{align}
Then putting all this together, we find
\beq
\alpha=\frac{\kappa^4 M^2}{16\pi^2(1-\bm  {\hat p_2\cdot\bm  \hat p_4})^2}
\eeq
\beq
\beta^{ij}=
    \Big( \frac{\kappa^4 }{16\pi^2(1 - \bm  {\hat p_2 \cdot \bm  \hat p_4} )^2}\Big)( \delta_{ij} +
   3   {\hat p_{2i}   \hat p_{2j}} )
\eeq
The term $\alpha$ gives exactly the limiting small momentum transfer gravitational deflection of a massless particles from a massive particle, \cite{Scadron}.
\section{Discussion and Conclusions}
The term proportional to $\alpha$, the lowest order monopole term, correspond to the scattering cross section of a single point like mass $M$, the analog of the Rutherford/Thompson cross section of a massless particle scattering from a point like Newtonian potential.  We see that the higher order terms coming from the $\beta^{ij}\delta_{ij}=\Big( \frac{\kappa^4 }{16\pi^2(1 - {\bm  {\hat \bm p_2}} \cdot \bm  {\bm {\bm {\bm {\hat p_4 }}}})^2 } \Big) 6$, contribute to the monopole.  This means that the scattering cross section is able to probe the non point like nature of the monopole part of the scattering particle's wave function.  This addition to the monopole contribution is given by
\beq
 \Big( \frac{\kappa^4 }{16\pi^2(1 - \bm  {\hat p_2} \cdot \bm  {\bm {\bm {\bm {\hat p_4 }}})}^2} \Big) 2 \lb\frac{3}{2 \sigma^2} - \frac{\vert\bm  r\vert^2}{\sigma^4} \frac{1}{1 + e^{\vert \bm  r \vert^2/\sigma^2}}  \rb .
\eeq

However the actual quadrupole-type contribution is interestingly nothing like what would be expected if the gravitational field behaved according to the Schrödinger-Newton prescription, \cite{Diosi,Penrose}.  The Schrödinger-Newton prescription would have the corresponding gravitational field as if one half the mass were concentrated at each of the two spatially nonlocal points, \cite{Diosi2}, a configuration which has a quadrupole moment
\beq
M\left(\frac{\vert \bm  r \vert^2}{3} \delta_{ij}-  r_{i}   r_{j}  \right) .
\eeq
Such a gravitational field yields a contribution to the (gravitational) scattering cross section
\beq
 \frac{\kappa^4M^2\omega_2^2}{16\pi^2(1-\bm  {\hat p_2\cdot\bm  \hat p_4})^2}(  \hat p_{4i}-\hat p_{2i})(  \hat p_{4j}-\hat p_{2j})\left(\frac{\vert \bm  r \vert^2}{3} \delta_{ij}-  r_{i}   r_{j}  \right)
\eeq
obtained from a presumed Newtonian potential scattering from two point sources of mass $M/2$ located at the two peaks of the spatially nonlocal wave function.  This result is not at all what we find from, Eqn.\eqref{11}
\beq
\frac{ \kappa^4 }{16\pi^2\sigma^4\lb 1 + e^{\vert \bm  r \vert^2/\sigma^2}\rb}  \frac{3{\hat p_{2i}   \hat p_{2j}}}{ {(1 - \bm  {\hat p_2 \cdot \bm  \hat p_4} )^2}}\left(\frac{\vert \bm  r \vert^2}{3} \delta_{ij}-  r_{i}   r_{j}  \right) ,
\eeq
where $\kappa=\sqrt{8\pi G}$ which gives a coefficient $4G^2$.  Indeed the quadrupole term from the calculation from one graviton exchange scattering is exponentially small as $\vert \bm  r\vert\gg \sigma$ compared to the result expected from scattering from a potential where the mass is split between two postions as prescribed by the Newton-Schrödinger formalism.   This is lethal to the Schrödinger-Newton formalism, the calculation corresponding to one graviton exchange is clearly more justifiable.   Additionally, our result shows that the wave function is only probed by the incoming massless particle's direction relative to the direction of separation $\bm  r$ something that is hopefully experimentally verifiable.  

Useful further calculations would be to compute the gravitational contribution to the self energy of a massive particle in a spatially nonlocal wave function.  One would look for the amplitude and behaviour of the self energy as a function of the spatial separation $r$ of the nonlocal wave function.  A behaviour as  $ 1/r$  of the self energy would correspond to the Newtonian potential and the corresponding $ 1/r^2$ law of attraction  of the two non-local lumps.  A calculation of the amplitude would indicate how the two nonlocal lumps behave gravitationally with respect to each other. 

It would be very interesting to measure any of these phenomena although probably not technically feasible presently.  However, it is not out of the realm of possibility to have a spatially non-local superposition about a $10^9$ atoms.  A Bose condensate would be proposed, and it can be imagined that such condensates could be launched in a atom interferometer.   Then the gravitational interactions of the two quantum superposed masses might be measurable \cite{K}.

\section*{Acknowledgments}
We thank the Indian Institute of Technology Bombay for its hospitality where this work was started. We thank the Ministère des relations internationales et de la francophonie du gouvernement du Québec and the Programme de Coopération Québec-Maharashtra and NSERC of Canada for financial support.  The work of VM was further supported by the Department of Physics and the Faculty of Graduate Studies of the Université de Montréal.

\bibliography{QGS}
\bibliographystyle{unsrt}

\end{document}